\documentstyle[aps,prb,multicol,epsf,graphics]{revtex}
\begin{document}
\draft
\title{Optimization of Gutzwiller Wavefunctions in Quantum Monte Carlo}
\author{Erik Koch,$^{(a,b)}$ Olle Gunnarsson,$^{(a)}$ 
        and Richard M.\ Martin$^{(b)}$}
\address{$^{(a)}$Max-Planck-Institut f\"ur Festk\"orperforschung, 
                D-70569 Stuttgart}
\address{$^{(b)}$Department of Physics, University of Illinois at 
                 Urbana-Champaign, Urbana, IL 61801}
\date{\today}
\maketitle
\begin{abstract}
Gutzwiller functions are popular variational wavefunctions for correlated
electrons in Hubbard models. Following the variational principle, we are
interested in the Gutzwiller parameters that minimize e.g.\ the expectation
value of the energy. Rewriting the expectation value as a rational function
in the Gutzwiller parameters, we find a very efficient way for performing
that minimization. The method can be used to optimize general Gutzwiller-type
wavefunctions both, in variational and in fixed-node diffusion Monte Carlo.
\end{abstract}
\pacs{71.10.Fd, 71.27.+a, 02.70.Lq}

\begin{multicols}{2}
\section{Introduction}

The Hubbard Hamiltonian\cite{Hubbard1,Hubbard2,Hubbard3} is a basic model
for studying correlated electrons. Despite its simple form it is very difficult
to solve, the main complication arising from the fact that the
interaction term is diagonal in real space, while the kinetic energy is simple
in momentum space. Except for special cases, namely the one\cite{LiebWu} and
the infinite-dimensional\cite{infrev} Hubbard model, no exact solutions are
known. For other dimensions we thus have to use approximate methods. The 
variational method has the advantage of being applicable over the whole
range of interaction strength. Naturally, the quality of variational 
calculations depends critically on the trial function and on our ability to
optimize its parameters.
A good trial function has to balance the opposing tendencies of the kinetic-
and the interaction term. 
Gutzwiller proposed such a wavefunction for the Hubbard model, 
the Gutzwiller wavefunction (GWF).\cite{Hubbard1,GWF2,GWF3} It introduces 
correlations into a Slater determinant by means of a correlation 
factor that is local in configuration space. Like the Jastrow factor in 
continuum-space wavefunctions,\cite{Jastrow} it works by reducing the weight
of configurations with large potential energy.
Unfortunately, also the GWF is hard to treat exactly.
Again, analytical results have only been obtained in 
one dimension,\cite{GWFd1,GWFd1a,GWFd1b,GWFd1c} and in 
infinite dimensions,\cite{GWFinf} where the Gutzwiller approximation (GA) 
becomes exact.\cite{GWF3,GArev} For $1<d<\infty$ we have to resort to 
numerical methods like variational Monte Carlo (VMC).\cite{Ceperley77,Horsch}
Another variational method that has been developed recently is fixed-node
diffusion Monte Carlo (FNDMC) for Fermions on a lattice.\cite{FNDMC,FNDMCa} 
Like variational Monte Carlo it uses a trial wavefunction, but the results 
are much closer to the exact ground state and depend much less on the trial 
wavefunction.
Since these Monte Carlo methods work in configuration space it is 
straightforward to implement trial functions with more correlation
factors. Such generalized Gutzwiller-type wavefunctions\cite{GWF3} that 
include for example correlations between empty and doubly occupied sites 
give improvements over the original Gutzwiller wavefunction.\cite{KaplanHorsch}

Our goal is to optimize the parameters in Gutzwiller-type wavefunctions in 
the framework of quantum Monte Carlo calculations. General methods for 
achieving this are  
correlated sampling\cite{Ceperley77,Umrigar88} or the recently developed 
stochastic gradient approximation.\cite{SGA}
Exploiting the particular form of Gutzwiller wavefunctions, we find a
different approach, which is equivalent to correlated sampling:
We observe that expectation values can be rewritten as the quotient
of two polynomials (i.e.\ a rational function) in the Gutzwiller parameters. 
Estimating the coefficients of these polynomials in a single Monte Carlo run, 
we can then easily minimize the energy expectation value by finding the minimum
of the corresponding rational function.
The implementation of this idea in variational Monte Carlo is described
in section \ref{VMC}. We show how the optimization works in practice and give 
some applications. In Sec.\ \ref{DMC} the optimization method is adapted to
work also in fixed-node diffusion Monte Carlo. This allows us to study the
effect of changing the fixed-node constraint in FNDMC.
Applications and results for wavefunctions with more parameters 
are given in Sec.\ \ref{appl}.

\section{Variational Monte Carlo}
\label{VMC}

To be specific, we consider the Hubbard model
\begin{equation}\label{HubbardHamil}
  H = -t\sum_{\langle i,j \rangle, \sigma} 
         c^{\dagger}_{i,\sigma} c^{\phantom\dagger}_{j,\sigma} 
      + U\sum_i n_{i,\uparrow} n_{i,\downarrow} ,
\end{equation}
where $c^{\dagger}_{i,\sigma}$ creates an electron with spin $\sigma$ on site 
$i$ and $n_{i,\sigma}=c^{\dagger}_{i,\sigma} c^{\phantom\dagger}_{i,\sigma}$. 
The sum in the kinetic term is over nearest-neighbor pairs, and the hopping
matrix element is used as the energy scale, i.e.\ $t\equiv1$.
The interaction term in (\ref{HubbardHamil}) may also be written as $U\,D$,
where $D=\sum_i n_{i,\uparrow} n_{i,\downarrow}$ is the operator that counts
the number of doubly occupied sites. The Gutzwiller wavefunction is then given
by
\begin{equation}\label{GWF}
  |\Psi(g)\rangle = g^D |\Phi_0\rangle ,
\end{equation}
\end{multicols}

\begin{figure*}
 \centerline{\epsfxsize=2.5in\epsffile{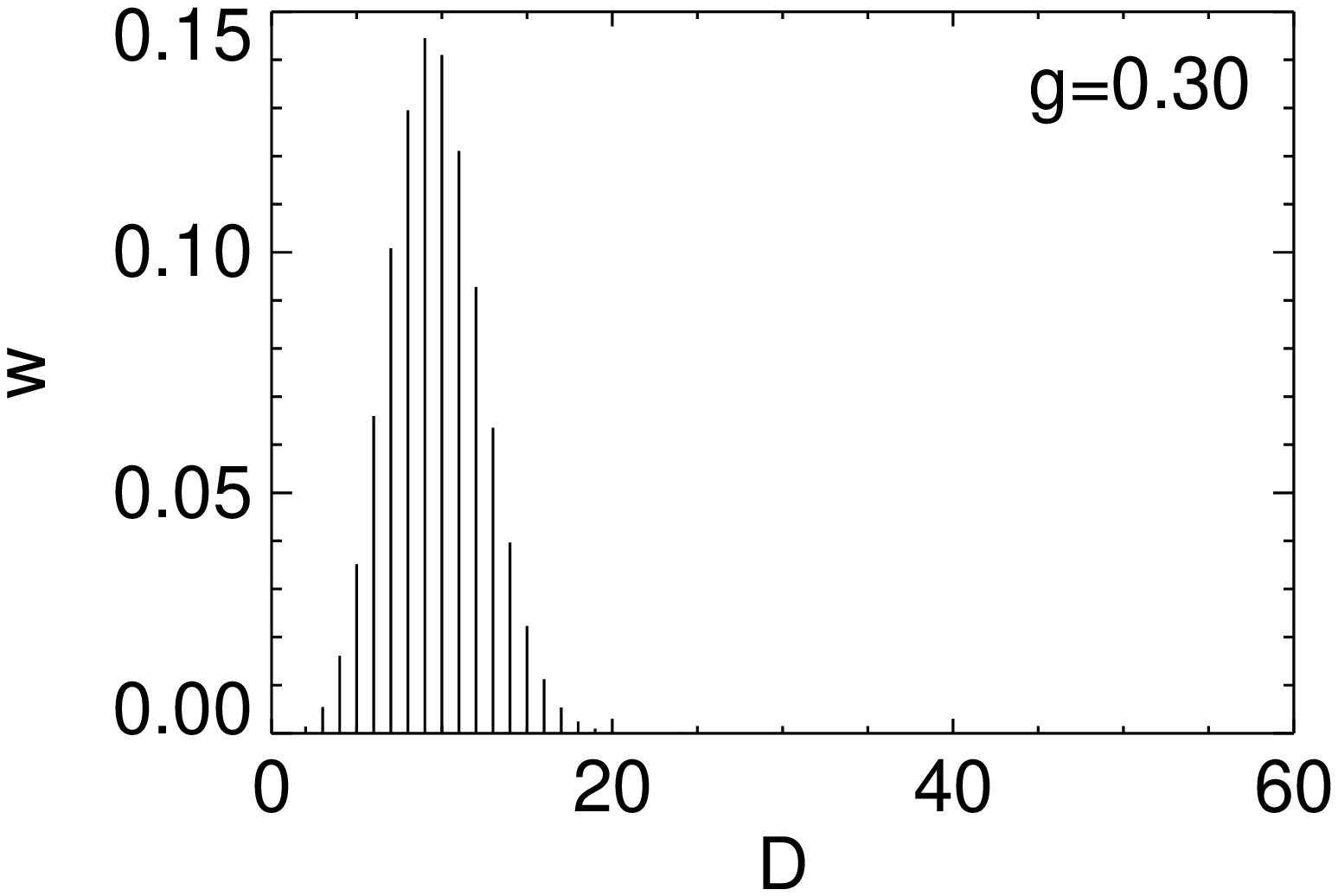}\hspace{-0.5cm}
             \epsfxsize=2.5in\epsffile{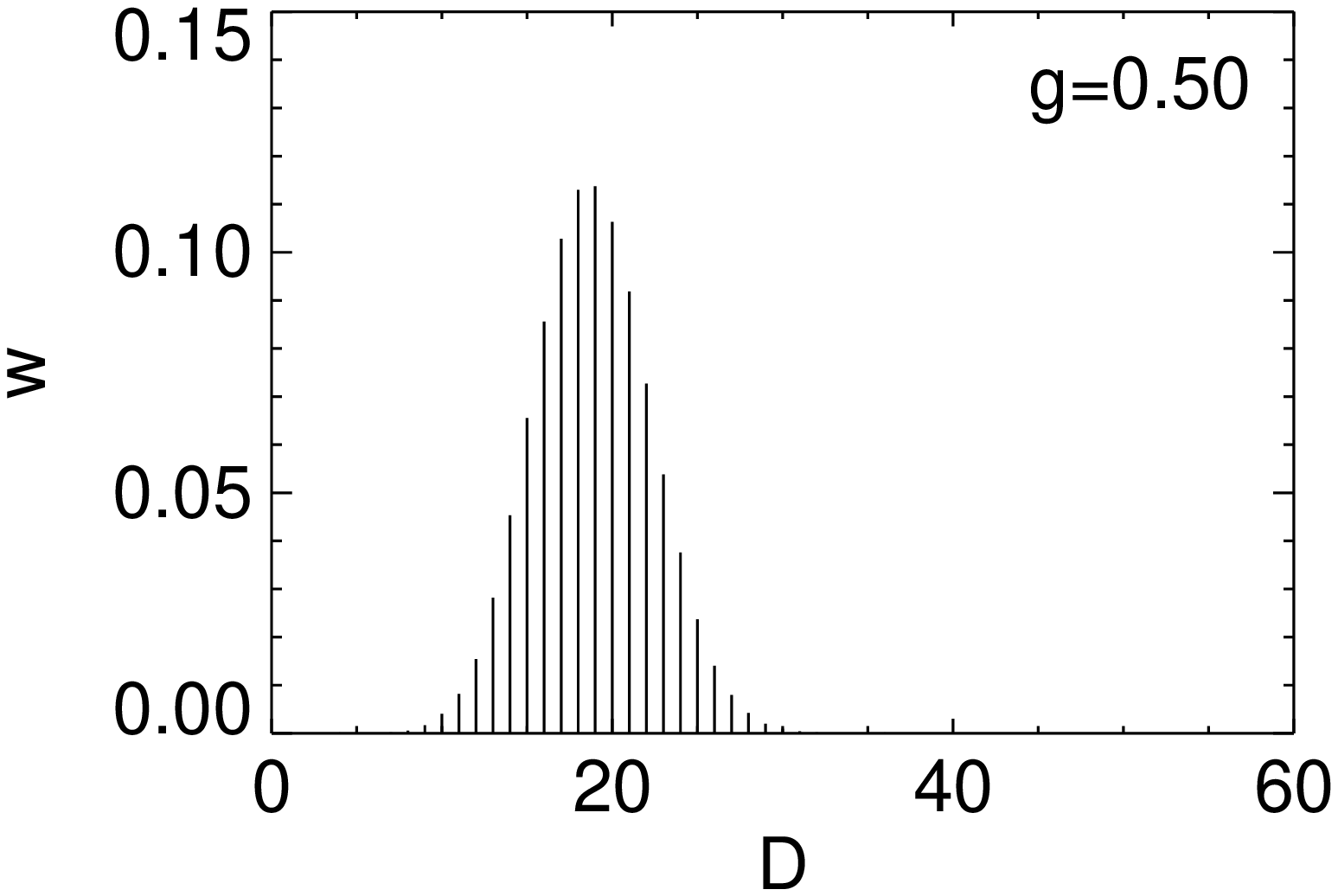}\hspace{-0.5cm}
             \epsfxsize=2.5in\epsffile{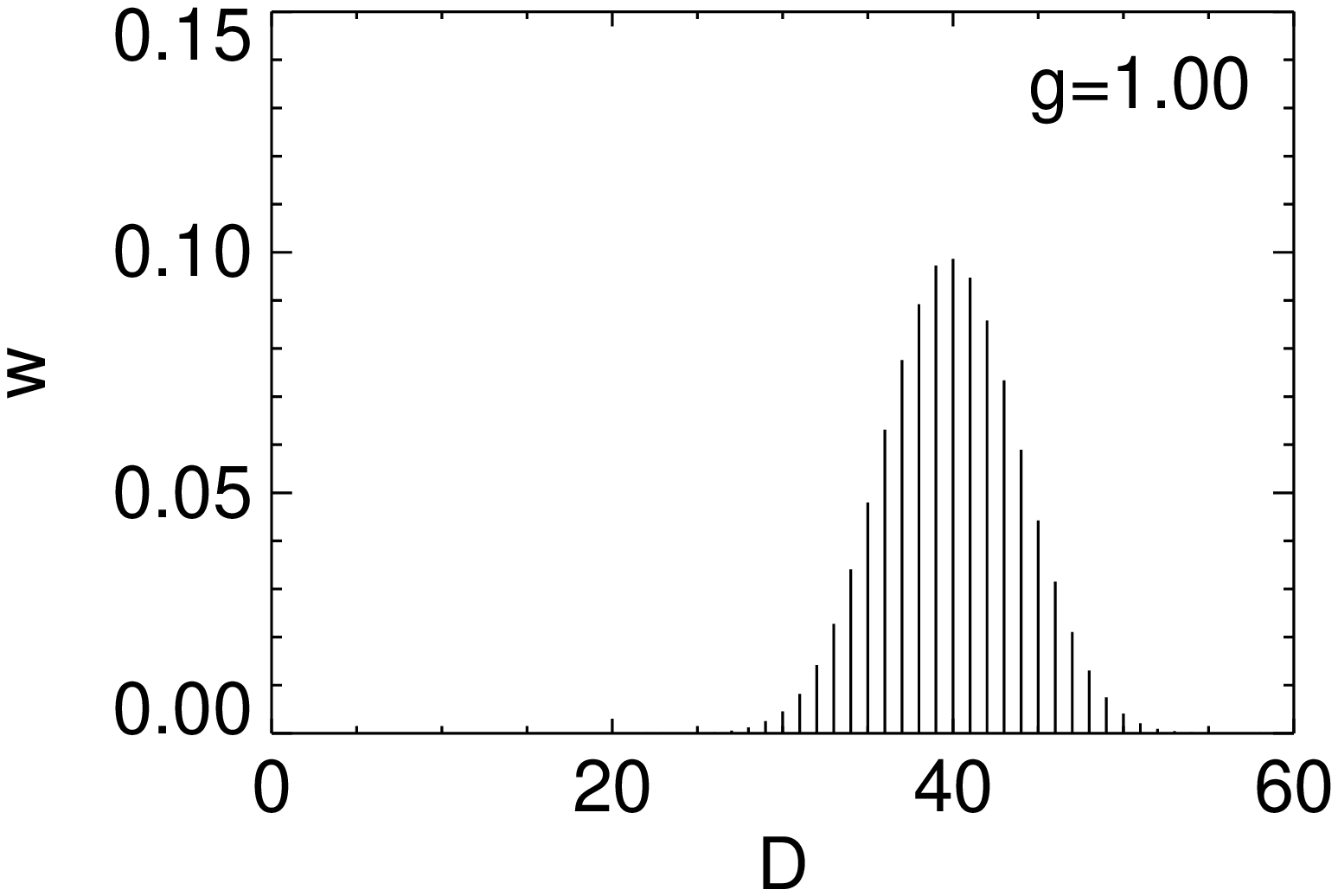}}
 \caption{\label{suppress}
   Weight of configurations with given number $D$ of double occupancies
   for the Gutzwiller wavefunction. (The weight $w(D)$ is defined as the 
   sum of $\Psi_T^2(R)$ over all configurations $R$ with $D$ doubly occupied
   sites.) Reducing the Gutzwiller factor $g$ 
   suppresses configurations $R$ for which the interaction energy 
   $E_{\rm int}(R)=U\,D(R)$ is large --- at the expense of increasing 
   the kinetic energy. The results shown are for $101+101$ 
   electrons on a $16\times16$ lattice.}
\end{figure*}

\begin{multicols}{2}
\noindent
where $0\le g \le 1$ is the Gutzwiller parameter and $|\Phi_0\rangle$ is the 
ground state wavefunction of the non-interacting system ($U=0$). The
Gutzwiller factor builds correlations into the Slater determinant by 
reducing the weight of configurations with large interaction energy, 
i.e.\ large number of doubly occupied sites. This is visualized in 
Fig.\ \ref{suppress}. 

In the following we briefly review the formalism of variational Monte Carlo 
for lattice systems and discuss a way to improve the efficiency of the 
simulation, especially for systems with strong correlation.
Then we introduce the method for optimizing the Gutzwiller parameter in a VMC
calculation.

\subsection{Modified Metropolis algorithm}

To calculate the energy expectation value for the Gutzwiller wavefunction
we have to perform a sum over all configurations $R$:
\begin{equation}\label{Evar}
 E_T = {\langle\Psi_T|H|\Psi_T\rangle \over \langle\Psi_T|\Psi_T\rangle}
     = {\sum_R E_{\rm loc}(R)\;\Psi_T^2(R) \over \sum_R \Psi_T^2(R)} ,
\end{equation}
where we have introduced the local energy for a configuration $R$:
\begin{eqnarray}
 E_{\rm loc}(R) 
 &=& \sum_{R'} {\langle\Psi_T|R'\rangle\,\langle R'|H|R\rangle
             \over \langle\Psi_T|R\rangle} \nonumber \\
 &=& -t \sum_{R'}\!'\;{\Psi_T(R')\over\Psi_T(R)} + U\,D(R) \label{Eloc} .
\end{eqnarray}
The prime in the last equation indicates that the sum is restricted to 
configurations $R'$ that are connected with $R$ by the Hamiltonian, i.e.\ 
configurations that can be reached from $R$ by hopping one electron to
a nearest-neighbor site. We call such configurations nearest-neighbors
in configuration space.

Since the number of configurations grows combinatorially with system-size, 
the sums in (\ref{Evar}) can in general only be performed for very small 
systems. For larger problems we can use Monte Carlo. 
The idea is to perform a random walk in the space of 
configurations, with transition probabilities $p(R\to R')$ chosen such
that the configurations $R_{\rm VMC}$ in the random walk have the probability
distribution function $\Psi_T^2(R)$. Then
\begin{equation}\label{Evmc}
 E_{\rm VMC} = 
  {\sum_{R_{\rm VMC}} E_{\rm loc}(R) \over \sum_{R_{\rm VMC}} 1}
 \approx E_T , 
\end{equation} 
where the last equality holds within statistical error-bars.
The transition probabilities are given by 
$p(R\to R')={1/N}\;\min[1,\Psi_T^2(R')/\Psi_T^2(R)]$, with
$N$ being the maximum number of possible transitions.\cite{Ntrans}
This choice fulfills detailed balance
\begin{equation}\label{detailedbalance}
  \Psi_T^2(R)\,p(R\to R') = \Psi_T^2(R')\,p(R'\to R) .
\end{equation}
For ergodicity it is sufficient to consider only transitions between
nearest neighboring configurations. The standard prescription is then to 
propose a transition $R\to R'$ with probability $1/N$ and accept it with 
probability $\min[1,\Psi_T^2(R')/\Psi_T^2(R)]$. This works well for $U$ not 
too large. In strongly correlated systems, however, the random walk will stay
for long times in configurations with a small number of double occupancies
$D(R)$, since most of the proposed moves will increase $D$ and hence be 
rejected with probability $\approx 1-g^{D(R')-D(R)}$.\cite{YoShi1}
There is, however, a way to integrate the time the walk stays in a given 
configuration out. To see how this works, we first observe that for the 
local energy (\ref{Eloc}) the ratio of the wavefunctions for all transitions
induced by the Hamiltonian have to be calculated. This in turn means that we
also know all transition probabilities $p(R\to R')$. We can therefore eliminate
any rejection (i.~e.\ make the acceptance ratio equal to one) by proposing
moves with probabilities
\begin{equation}
  \tilde{p}(R\to R') = {p(R\to R')\over\sum_{R'} p(R\to R')} 
                     = {p(R\to R')\over 1-p_{\rm stay}(R)} .
\end{equation}
Checking detailed balance (\ref{detailedbalance}) we find that now we are
sampling configurations $\bar{R}_{VMC}$ from the probability distribution
function $\Psi_T^2(R)\,(1-p_{\rm stay}(R))$. To compensate for this we assign
a weight $w(R)=1/(1-p_{\rm stay}(R))$ to each configuration $R$. The energy 
expectation value is then, within statistical error-bars, given by
\begin{equation}\label{Evmcwght}
 E_T \approx 
 {\sum_{\bar{R}_{VMC}} w(\bar{R})\,E_{\rm loc}(\bar{R}) \over
  \sum_{\bar{R}_{VMC}} w(\bar{R})} .
\end{equation} 
The above method is quite efficient since it ensures that in every Monte Carlo
step a new configuration is visited. In other words, the acceptance ratio is
one. Instead of staying in a configuration 
where $\Psi_T$ is large, this configuration is weighted with the expectation
value of the number of times the simple Metropolis algorithm would stay there.
This is particularly convenient for simulations of systems with strong 
correlations: Instead of having to do longer and longer runs as $U$ is 
increased, the above method produces, for a fixed number of Monte Carlo 
steps, results with comparable error-bars.

\subsection{Optimization}

We now turn to the problem of minimizing the energy expectation value 
(\ref{Evar}) as a function of the variational parameters in the trial function.
To this end we could simply perform independent VMC calculations for a set 
of different parameters. It is, however, difficult to compare the energies 
from independent calculations since each VMC result comes with its own 
statistical errors. This problem can be avoided with 
correlated sampling.\cite{Ceperley77,Umrigar88} The idea is to use the same 
random walk in calculating the expectation value for different trial functions.
This reduces the relative errors and hence makes it easier to find the minimum.

Let us assume then that we have generated a random walk $\{R_{\rm VMC}\}$ 
for the trial function $\Psi_T$. Using the {\em same} random walk, we can
also estimate the energy expectation value (\ref{Evmc}) for a different trial 
function $\tilde{\Psi}_T$. To do so we have to compensate for the fact that 
the configurations have the probability distribution $\Psi^2_T$ instead of 
$\tilde{\Psi}_T^2$ by introducing reweighting factors 
\begin{equation}\label{corrsmpl}
 \tilde{E}_T \approx
  {\sum_{R_{\rm VMC}} \tilde{E}_{\rm loc}(R)\;\tilde{\Psi}_T^2(R)/\Psi_T^2(R) 
   \over
   \sum_{R_{\rm VMC}}                         \tilde{\Psi}_T^2(R)/\Psi_T^2(R)} .
\end{equation}
Likewise, (\ref{Evmcwght}) is reweighted into
\begin{equation}
 \tilde{E}_T \approx
  {\sum_{\bar{R}_{\rm VMC}} w(\bar{R})\,\tilde{E}_{\rm loc}(\bar{R})\;
                            \tilde{\Psi}_T^2(\bar{R})/\Psi_T^2(\bar{R}) 
   \over
   \sum_{\bar{R}_{\rm VMC}} w(\bar{R})\,                        
                            \tilde{\Psi}_T^2(\bar{R})/\Psi_T^2(\bar{R})} .
\end{equation}
Also the local energy $\tilde{E}_{\rm loc}(R)$ can be rewritten to contain the
new trial function only in ratios with the old one.
For Gutzwiller functions this implies a drastic simplification. Since they 
differ only in the Gutzwiller factor, the Slater determinants cancel, leaving
only powers $(\tilde{g}/g)^{D(R)}$:
\begin{equation}\label{gcorr}
 E_T(\tilde{g}) \approx 
  {\sum_{R_{\rm VMC}} \tilde{E}_{\rm loc}(R)\;(\tilde{g}/g)^{2\,D(R)} \over
   \sum_{R_{\rm VMC}}                         (\tilde{g}/g)^{2\,D(R)}      }
\end{equation}
\begin{figure}
 \centerline{\epsfxsize=2.7in \epsffile{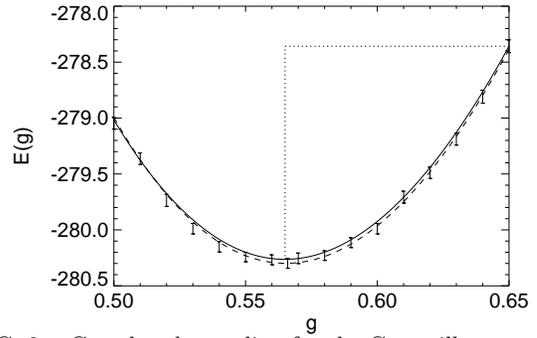}}
 \caption{\label{VMCcorrsmpl}
  Correlated sampling for the Gutzwiller parameter $g$. The results
  shown are for $101+101$ electrons on a $16\times16$ lattice and $U=4$.
  The full curve shows $E_T(g)$ calculated using $g=0.65$ as Gutzwiller 
  parameter. The predicted minimum $g_{\rm min}$ is indicated by the dotted 
  line. The dashed line gives the correlated sampling curve obtained from a
  calculation with $g_{\rm min}=0.566$. Both calculations find the same 
  minimum. The error-bars show the results of independent VMC runs for 
  different Gutzwiller parameters. We can clearly see the statistical 
  fluctuations in the results of the separate VMC runs, which, of course, 
  are absent in the correlated sampling curves.
 }
\end{figure}

\noindent
and 
\begin{equation}\label{gloc}
 \tilde{E}_{\rm loc}(R) = 
 -t\sum_{R'}\!'\, (\tilde{g}/g)^{(D(R')-D(R)}\;{\Psi_T(R')\over\Psi_T(R)} 
 + U\,D(R) .
\end{equation}
Since the number of doubly occupied sites $D(R)$ for a configuration $R$ is an
{\em integer}, we can then rearrange the sums in (\ref{gcorr}) and (\ref{gloc})
into polynomials in $\tilde{g}/g$. The energy expectation value for any 
Gutzwiller parameter $\tilde{g}$ is then given by a rational function in the 
variable $\tilde{g}/g$, where the coefficients only depend on the fixed trial
function $|\Psi(g)\rangle$.

It is then clear how we proceed to optimize the Gutzwiller parameter in 
variational Monte Carlo. We first pick a reasonable $g$ and perform a VMC run 
for $|\Psi(g)\rangle$ during which we also estimate the coefficients of the 
above polynomials. We can then easily calculate $E_T(\tilde{g})$ by evaluating 
the rational function in $\tilde{g}/g$. Since there are typically only of the
order of a few tens non-vanishing coefficients (cf.\ the distribution of 
weights shown in Fig.\ \ref{suppress}), this is a very efficient process. 

Figure \ref{VMCcorrsmpl} shows how the method works in practice.
Although we deliberately picked a bad starting point we still find the right
minimum. The correlated sampling curves even seem to coincide within the
statistical errors. This is, however, not true for the whole range of 
Gutzwiller parameters. When $\tilde{g}$ differs too much from $g$, the method
breaks down.  To understand this we again turn to Fig.\ \ref{suppress}. We see 
that most configurations in a random walk generated with, say, $g=0.50$ will 
have about $20$ doubly occupied sites. In the Monte Carlo run we therefore 
sample the coefficients for $(\tilde{g}/g)^{2\times20}$ best while the 
statistics for much larger or smaller powers is poor. But it is exactly these
poorly sampled coefficients that we need in calculating the energy expectation
value for trial functions with $\tilde{g}$ much different from $g$. We can thus
use the overlap of the wavefunctions $\langle\Psi(\tilde{g})|\Psi(g)\rangle$ as
a measure for the reliability of the calculated energy $E_T(\tilde{g})$. 
Like the energy expectation value itself, it can be recast in the form
of a rational function, the coefficients of which can be sampled during
the VMC run: 
\begin{eqnarray}
 \langle\Psi(\tilde{g})|\Psi(g)\rangle
 &=& {\sum_R \tilde{\Psi}(R)\,\Psi(R) \over
      \sqrt{\sum_R\tilde{\Psi}^2(R)\;\sum_R\Psi^2(R)}}\nonumber\\
 &=& {\sum_{R_{\rm VMC}} (\tilde{g}/g)^{D(R)} \over
      \sqrt{\sum_{R_{\rm VMC}} (\tilde{g}/g)^{2D(R)}\;\sum_{R_{\rm VMC}} 1}} .
\end{eqnarray}
To get a feeling for the overlap as a function of $\tilde{g}$ for fixed $g$,
we can make a rough estimate using the Gutzwiller approximation. Expanding 
around $g$ we find that the overlap looks like a Gaussian: 
$\exp[-M\,(\tilde{g}-g)^2/\sigma_0^2]$, with M the number of lattice sites.
As expected, for $g$ and $\tilde{g}$ fixed, the overlap goes to zero 
exponentially with system size. $\sigma_0$ is a function of $g$ and the filling.
It generally decreases with $g$. This can be understood by looking at 
Fig.\ \ref{suppress} as for small $g$ the weights are peaked more sharply 
than for larger Gutzwiller parameters. For half filling the width of the 
Gaussian is given by $\sigma_0=\sqrt{2g}\;2(1+g)$.
 
\begin{figure}[t]
 \centerline{\epsfxsize=3.2in \epsffile{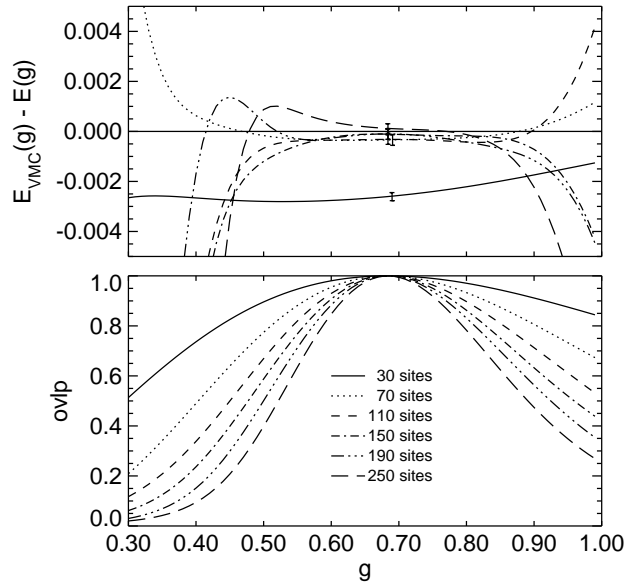}}
 \vspace{1ex}
 \caption[]{\label{Metzner}
   Comparison of the correlated sampling results for finite Hubbard chains
   to the exact result for the infinite chain with $U=2$. The energies from
   the VMC calculations are indicated by the error-bars in the upper figure.
   The curves give the corresponding results $E_{\rm VMC}(g)$ from evaluating 
   expression (\ref{gcorr}). Clearly there are finite-size errors, especially
   for the small systems, but apart from that the agreement between the
   correlated sampling and the exact energy $E(g)$ for the infinite system
   is remarkable, even for fairly small overlaps of the trial functions, as
   can be seen from the lower panel. Eventually, however, when the overlaps
   become too small (here $\lesssim 0.4$) correlated sampling breaks down.
  }
\end{figure}

The relation between the overlap and the reliability of the expression 
(\ref{gcorr}) can best be seen by comparing VMC calculations to exact
energies $E(g)$. An example is shown in Fig.\ \ref{Metzner}. There we compare
the results of VMC calculations for finite Hubbard chains of different size
with the exact result for the infinite chain.\cite{GWFd1,GWFd1a} Clearly there
are systematic errors in the energy coming from finite size effects in the VMC 
calculations. But apart from that we find a remarkable agreement between the 
exact energy $E(g)$ and the result of the correlated sampling, even for fairly 
small overlaps. It is also evident from the figure that the overlap for given 
$g$ and $\tilde{g}$ decreases with system size, making optimizations more and 
more difficult, the larger the system. 

We finally mention some straightforward modifications of the scheme we have 
described
above. There are situations where it is more appropriate to minimize the 
variance in the local energy $\sigma^2(g)$ rather than the energy 
$E(g)$.\cite{Umrigar88} Since the variance can also be rewritten in terms of
a rational function in $\tilde{g}/g$, variance optimization can be implemented 
in much the same way as the energy minimization that we have described here.
Furthermore, it is clear that the method is not restricted to the plain
Gutzwiller wavefunction but can be generalized to trial functions with more 
correlation factors of the type $r^{c(R)}$. As long as the correlation function
$c(R)$ is integer-valued on the space of configurations, expectation values for
such trial functions can still be rewritten
\phantom{as rational functions.}

\vspace{-4ex}
\noindent
as rational functions. The only 
difference to the simpler case described above is that now the rational function
is multivariate, reflecting the fact that there is more than one variational
parameter.

\section{Fixed-Node Diffusion Monte Carlo}
\label{DMC}

We now turn to the optimization of the trial function in fixed-node diffusion
Monte Carlo (FNDMC). In this method the ground state wavefunction of a given
Hamiltonian is projected out from a trial function by repeated application of
a suitable operator. For Fermion systems this approach is plagued by the 
infamous sign-problem, which causes an exponential decay of the signal-to-noise
ratio during a simulation. One approach to evade this problem is to use a 
trial function to fix the nodes of the wavefunctions. This is the 
fixed-node approximation, which gives variational estimates of the ground state
energy as a function of the trial function. Clearly a change in the Jastrow 
factor, will not change the nodes of the trial function. Hence one would expect
the FNDMC results to be independent of such changes. This is actually true for
systems in continuum-space.\cite{CeperleyAlder,CeperleyAlderII} For lattice 
problems, the fixed-node approximation is somewhat more subtle, and, as it 
turns out, results also depend on the Gutzwiller parameters.\cite{FNDMC,FNDMCa}
Since fixed-node diffusion Monte Carlo, like VMC, is a variational method, it 
is then important to optimize the Gutzwiller factors. This optimization can 
again be done using correlated sampling. As in the preceeding section, 
we can rewrite the expression for the energy as a function of the Gutzwiller
parameter in terms of a ratio of polynomials. Now, however, the order of the 
polynomials is not a fixed, small number but increases with the number of Monte
Carlo steps. It is therefore not feasible to sample the coefficients directly, 
as we did above.

We will first briefly review the aspects of fixed-node diffusion Monte Carlo
that are important for implementing an optimization scheme for Gutzwiller
parameters, then describe how to perform the optimization, and finally show 
how the method works in practice. For more in-depth discussions and some 
applications of the FNDMC method see, e.g., Refs.\ 
\onlinecite{FNDMC,FNDMCa,Buonaura98,BacheletNATO}.

\subsection{Fixed-node approximation}

Diffusion Monte Carlo \cite{GFMC} allows us, in principle, to sample the true 
ground state of a Hamiltonian $H$. The basic idea is to use a projection
operator which has the lowest eigenstate as a fixed point. For a lattice 
problem, where the spectrum is bounded $E_n\in[E_0,E_{\rm max}]$, the 
projection is given by
\begin{equation}\label{proj}
  |\Psi^{(n+1)}\rangle = [1-\tau(H-E_0)]\;|\Psi^{(n)}\rangle ,
\end{equation}
starting with $|\Psi^{(0)}\rangle=|\Psi_T\rangle$.
If $\tau<2/(E_{\rm max}-E_0)$ and $|\Psi_T\rangle$ has a non-vanishing overlap
with the ground state, the above iteration converges to $|\Psi_0\rangle$. There
is no time-step error involved.
Because of the prohibitively large dimension of the many-body Hilbert space,
the matrix-vector product in (\ref{proj}) cannot be done exactly. Instead, we
rewrite the equation in configuration space
\begin{eqnarray}\label{iter}
  &&\sum |R'\rangle\langle R'|\Psi^{(n+1)}\rangle = \\
  &&\quad \sum_{R,R'} |R'\rangle
     \underbrace{\langle R'|1-\tau(H-E_0)|R\rangle}_{\equiv F(R',R)}
       \langle R|\Psi^{(n)}\rangle \nonumber
\end{eqnarray}
and perform the propagation in a stochastic sense: $\Psi^{(n)}$ is
represented by an ensemble of configurations $R$ with weights $w(R)$.
The transition matrix element $F(R',R)$ is rewritten as a transition
probability $p(R\to R')$ times a normalization factor $m(R',R)$. The iteration
(\ref{iter}) is then performed stochastically as follows: For each $R$ we pick
a new configuration $R'$ with probability $p(R\to R')$ and multiply its weight
by $m(R',R)$. Then the new ensemble of configurations $R'$ with their
respective weights represents $\Psi^{(n+1)}$. Importance sampling
decisively improves the efficiency of this process by replacing $F(R',R)$ with
$G(R',R)=\langle\Psi_T|R'\rangle\,F(R',R)/\langle R|\Psi_T\rangle$ so
that transitions from configurations where the trial function is small
to configurations with large trial function are enhanced.
(\ref{iter}) then takes the form
\begin{eqnarray*}
  &&\sum |R'\rangle\langle\Psi_T| R'\rangle\langle R'|\Psi^{(n+1)}\rangle = \\
  &&\quad \sum_{R,R'} |R'\rangle\,G(R',R)\,
               \langle\Psi_T|R\rangle\,\langle R|\Psi^{(n)}\rangle 
\end{eqnarray*}
i.e.\ the ensemble of configurations now represents the product 
$\Psi_T\,\Psi^{(n)}$,
and the transition probabilities are given by $p(R\to R')=|G(R',R)|/m(R',R)$
with $m(R',R)={\rm sign}[G(R',R)]\,\sum_{R''} |G(R'',R)|$ absorbing the sign 
of $G(R',R)$ and ensuring normalization. After a large number $n$ of iterations
the ground state energy is then given by the mixed estimator
\begin{equation}\label{mixedest}
  E_0^{(n)}
  = {\langle\Psi_T|H|\Psi^{(n)}\rangle \over \langle\Psi_T|\Psi^{(n)}\rangle}
  \approx {\sum_R E_{\rm loc}(R)\;w^{(n)}(R) \over \sum_R w^{(n)}(R)}
\end{equation}
with $w^{(n)}(R_n)=\prod_{i=1}^n m(R_i,R_{i-1})$.
As long as the evolution operator has only non-negative matrix elements
$G(R',R)$, all weights $w(R)$ will be positive. If, however, $G$ has
negative matrix elements there will be both configurations with positive and
negative weight. Their contributions to the estimator (\ref{mixedest})
tend to cancel so that eventually the statistical error dominates, rendering
the simulation useless. This is the infamous sign problem.
A straightforward way to get rid of the sign problem is to remove the
offending matrix elements from the Hamiltonian, thus defining a new Hamiltonian
$H_{\rm eff}$ by
\begin{equation}\label{offdiag}
   \langle R'|H_{\rm eff}| R\rangle = \left\{
     \begin{array}{cc}
                 0           & \mbox{ if $G(R',R)<0$} \\
      \langle R'|H| R\rangle & \mbox{ else}
     \end{array}\right. .
\end{equation}
For each off-diagonal element $\langle R'|H| R\rangle$ that has been removed,
a term is added to the diagonal:
\begin{displaymath}
  \langle R|H_{\rm eff}|R\rangle
       = \langle R|H|R\rangle
       + \sum_{R'\;sf} \Psi_T(R')\langle R'|H|R\rangle/\Psi_T(R) .
\end{displaymath}
This is the fixed-node approximation for lattice Hamiltonians introduced 
in Ref.~\onlinecite{FNDMC}. $H_{\rm eff}$ is by construction free of the
sign problem and variational, i.e.\ $E_0^{\rm eff}\ge E_0$, and if
if $\Psi_T(R')/\Psi_T(R)=\Psi_0(R')/\Psi_0(R)$ for all $R$, $R'$ with 
$G(R',R)<0$ the method is exact.

Fixed-node diffusion Monte Carlo for a lattice Hamiltonian thus means that
we choose a trial function from which we construct an effective Hamiltonian,
the ground state of which is determined by diffusion Monte Carlo. From the 
equations defining $H_{\rm eff}$ we can then understand how the fixed-node 
results depend on the trial function. Clearly, the off diagonal elements 
(\ref{offdiag}) only depend on the sign of the trial function. Since the 
Gutzwiller term is just a non-negative prefactor, a change of $g$ will not 
affect these matrix elements. On the other hand, the diagonal elements 
$\langle R|H_{\rm eff}|R\rangle$ can contain ratios of the trial function on 
neighboring configurations. In many cases these configurations will differ
in the number of doubly occupied sites. Then the Gutzwiller terms will not 
cancel and the diagonal element of the effective Hamiltonian will depend 
on the Gutzwiller factor.

\subsection{Correlated sampling}

We are now in the position to describe how we can optimize $\Psi_T$ or, 
equivalently, optimize $H_{\rm eff}$ using correlated sampling. The idea is
again to calculate the energy for a modified Hamiltonian $\tilde{H}_{\rm eff}$ 
using a random walk generated for the original Hamiltonian $H_{\rm eff}$.
To find the reweighting factors involved in the calculation we rewrite the
mixed estimator 
\begin{eqnarray}
  \tilde{E}_0^{(n)}
  &=&{\langle\tilde{\Psi}_T|\tilde{H}_{\rm eff}|\tilde{\Psi}^{(n)}\rangle
    \over \langle\tilde{\Psi}_T|\tilde{\Psi}^{(n)}\rangle} \nonumber \\
  &=& {\sum \tilde{E}_{\rm loc}(R)\,
         \prod_{i=1}^n \tilde{G}(R_i,R_{i-1})\,\tilde{\Psi}_T^2(R_0)
       \over \sum \prod_{i=1}^n \tilde{G}(R_i,R_{i-1})\,\tilde{\Psi}_T^2(R_0)} 
      \label{tildemixed} .
\end{eqnarray}
At first glance it looks like the local energy $\tilde{E}_{\rm loc}(R)$ has 
to be calculated for $\tilde{H}_{\rm eff}$, but keeping track of the sign-flip
terms in the diagonal of the effective Hamiltonian we find
\begin{equation}
  \tilde{E}_{\rm loc}(R) = \sum_{R'} {\tilde{\Psi}_T(R')\over\tilde{\Psi}_T(R)}
                                     \langle R'|H|R\rangle ,
\end{equation}
which is just the local energy for the original Hamiltonian. Hence, the local
energy can be dealt with as in variational Monte Carlo, cf.\ eqn.\ (\ref{gloc}).

Correlated sampling now means that in the expression (\ref{tildemixed}) for the 
mixed estimator we do Monte Carlo for the terms
$\prod_{i=1}^n G(R_i,R_{i-1})\,\Psi_T^2(R_0)$ instead of 
$\prod_{i=1}^n \tilde{G}(R_i,R_{i-1})\,\tilde{\Psi}_T^2(R_0)$. The reweighting
factors are thus given by
\begin{equation}\label{reweightdmc}
  \prod_{i=1}^n {\tilde{G}(R_i,R_{i-1})\over G(R_i,R_{i-1})}\;
                {\tilde{\Psi}_T^2(R_0) \over\Psi_T^2(R_0)} .
\end{equation}

Reintroducing the plain (not importance-sampled) projection operator 
$F=1-\tau(H-E_0)$, we can rewrite the reweighting factor for a simple 
Gutzwiller wavefunction into
\begin{equation}
  \prod_{i=1}^n {\tilde{F}(R_i,R_{i-1})\over F(R_i,R_{i-1})}\;
                \left({\tilde{g}\over g}\right)^{D(R_0)+D(R_n)} 
\end{equation}
with the ratio $\tilde{F}(R',R)/F(R',R)=1$ for $R'\ne R$, and
\begin{equation}\label{rewf}
 {\tilde{F}(R,R)\over F(R,R)} =
 {1-\tau\left[\sum\limits_{R'\;sf} {\Psi_T(R')\over\Psi_T(R)}
          \langle R'|H|R\rangle \left({\tilde{g}\over g}\right)^{\Delta(R',R)}
         \hspace{-1.5ex}-E_0\right]
  \over
  1-\tau\left[\sum\limits_{R'\;sf} {\Psi_T(R')\over\Psi_T(R)}
                   \langle R'|H|R\rangle  - E_0\right]
 }
\end{equation}
with $\Delta(R',R)=D(R')-D(R)$. This expression is again (as in variational
Monte Carlo) a low order polynomial in $\tilde{g}/g$. The reweighting factor
is, however, a product of \phantom{many such polynomials}

\vspace{-4ex} 
\begin{figure}[t]
 \centerline{\epsfxsize=3.1in \epsffile{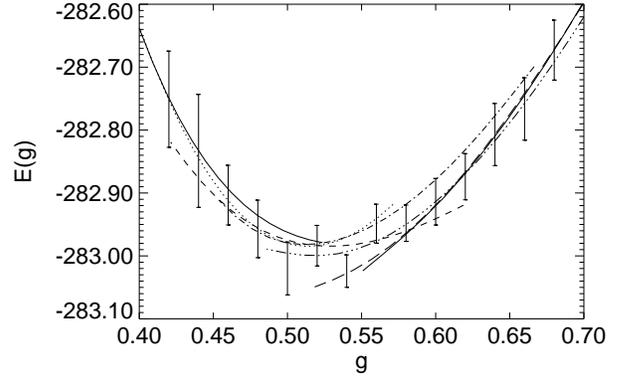}}
 \vspace{1ex}
 \caption[]{\label{dmcs}
  Correlated sampling for the Gutzwiller parameter $g$ in fixed-node diffusion
  Monte Carlo. The results are for the same system as in Fig.\ 
  \ref{VMCcorrsmpl} ($101+101$ electrons on a $16\times16$ lattice, $U=4$).
  The error-bars show the results of independent FNDMC runs for different
  Gutzwiller parameters. The lines give the results of correlated sampling
  for trial functions with 
  $g=0.44$, 0.48, 0.52, 0.56, 0.60, 0.64, and 0.68.
  Note that in fixed-node diffusion Monte Carlo the energy depends much less
  on the trial function than in variational Monte Carlo: The energy scale is 
  reduced by a factor 5 compared to Fig.\ \ref{VMCcorrsmpl}, although the 
  range of $g$-values is expanded by a factor of two.
  }
\end{figure}

\noindent
many such polynomials, therefore the order of the 
polynomial representing (\ref{reweightdmc}) increases with the number $n$ of
Monte Carlo steps. It is therefore not practical to directly estimate the
ever increasing number of coefficients for the reweighting factor. But since 
we still can easily calculate the coefficients for the factors (\ref{rewf})
we may use them to evaluate the mixed estimator $E_0^{(n)}(\tilde{g})$ in each
iteration on a set of {\em predefined} values $\tilde{g}_i$ of the Gutzwiller
parameter.

To see how the method works in practice we look at the same system as in
Fig.\ \ref{VMCcorrsmpl}: $101+101$ electrons on a $16\times16$ lattice, with
$U=4$. Figure \ref{dmcs} shows the result of independent FNDMC runs and the
corresponding correlated sampling results. We see that our method gives good 
predictions of the optimum Gutzwiller parameter, even for runs with trial
functions that are quite far from optimum. Comparing with variational Monte 
Carlo, we first notice that the variational energy in FNDMC is substantially 
lower (by $\approx 2.5\;eV$). We furthermore see that in the fixed-node method
there is a pronounced minimum in the energy as a function of the Gutzwiller
parameter, although the energy dependence is much smaller than in variational
Monte Carlo. This is to be expected since in FNDMC only certain features of
the trial function (namely the ratio of the trial function across the nodes)
enter the calculation. The situation is, however, different from the fixed-node
method for wavefunctions defined in continuum-space, for which changing the
Jastrow factor does not change the position of the nodal surfaces and hence 
also does not systematically change the result of the fixed-node calculation.
We finally notice that the optimum Gutzwiller parameter for FNDMC is slightly
smaller than that obtained from VMC.

\section{Applications}
\label{appl}

We have made practical use of the method for optimizing Gutzwiller parameters
in our investigations of the doped Fullerides. In this context we work with
a Hubbard-like Hamiltonian that describes the conduction electrons in the
three-fold degenerate $t_{1u}$ band:
\begin{equation}\label{C60Hamil}
  H=\sum_{\langle ij\rangle} \sum_{nn'\sigma} t_{in,jn'}\;
              c^\dagger_{in\sigma} c^{\phantom{\dagger}}_{jn'\sigma}
 +\;U\sum_i\hspace{-0.7ex} \sum_{(n\sigma)<(n'\sigma')}\hspace{-1.5ex}
       n_{i n\sigma} n_{i n'\sigma'} ,
\end{equation}
where $t_{in,jn'}$ is the hopping integral between orbital $n$ of the molecule 
on site $i$ and orbital $n'$ of molecule $j$, and $U$ is the Coulomb 
interaction for electrons on the same molecule.  For more details on the 
underlying physics see Refs.\ \onlinecite{rmpc60,mott,susc,screenKirchb,NATO}.
In the Monte Carlo calculations for the above Hamiltonian, we use trial 
functions of the Gutzwiller-type
\begin{equation}\label{GWFC60}
  |\Psi_T(U_0,g)\rangle = g^D |\Phi(U_0)\rangle ,
\end{equation}
where besides the Gutzwiller parameter $g$ we also use different types of
Slater determinants $\Phi$.

\subsection{More Gutzwiller parameters}

\begin{figure}[t]
  \centerline{\resizebox{3.2in}{!}{\rotatebox{270}{\includegraphics{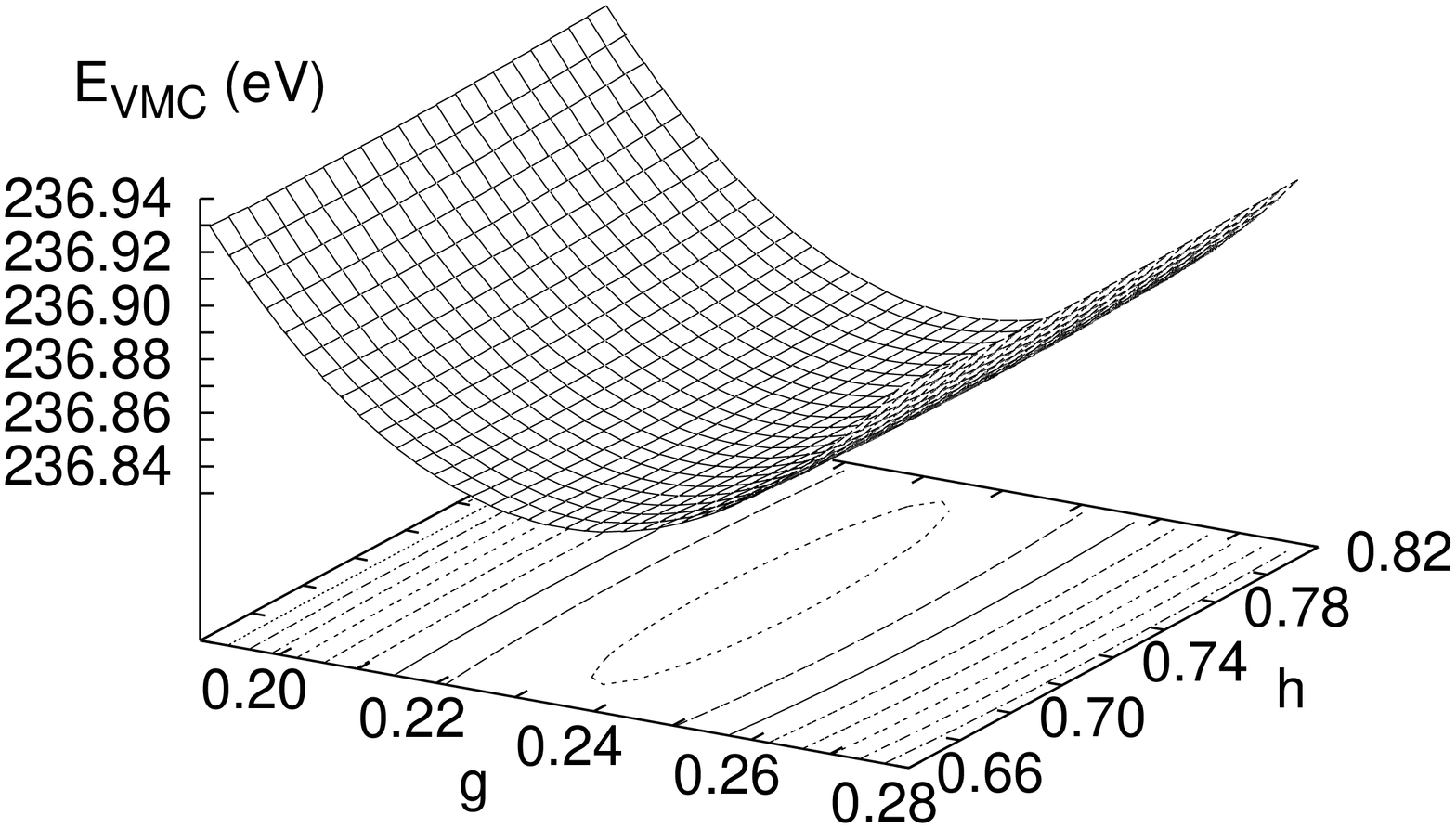}}}}
  \vspace{-5ex}
  \centerline{\resizebox{3.2in}{!}{\rotatebox{270}{\includegraphics{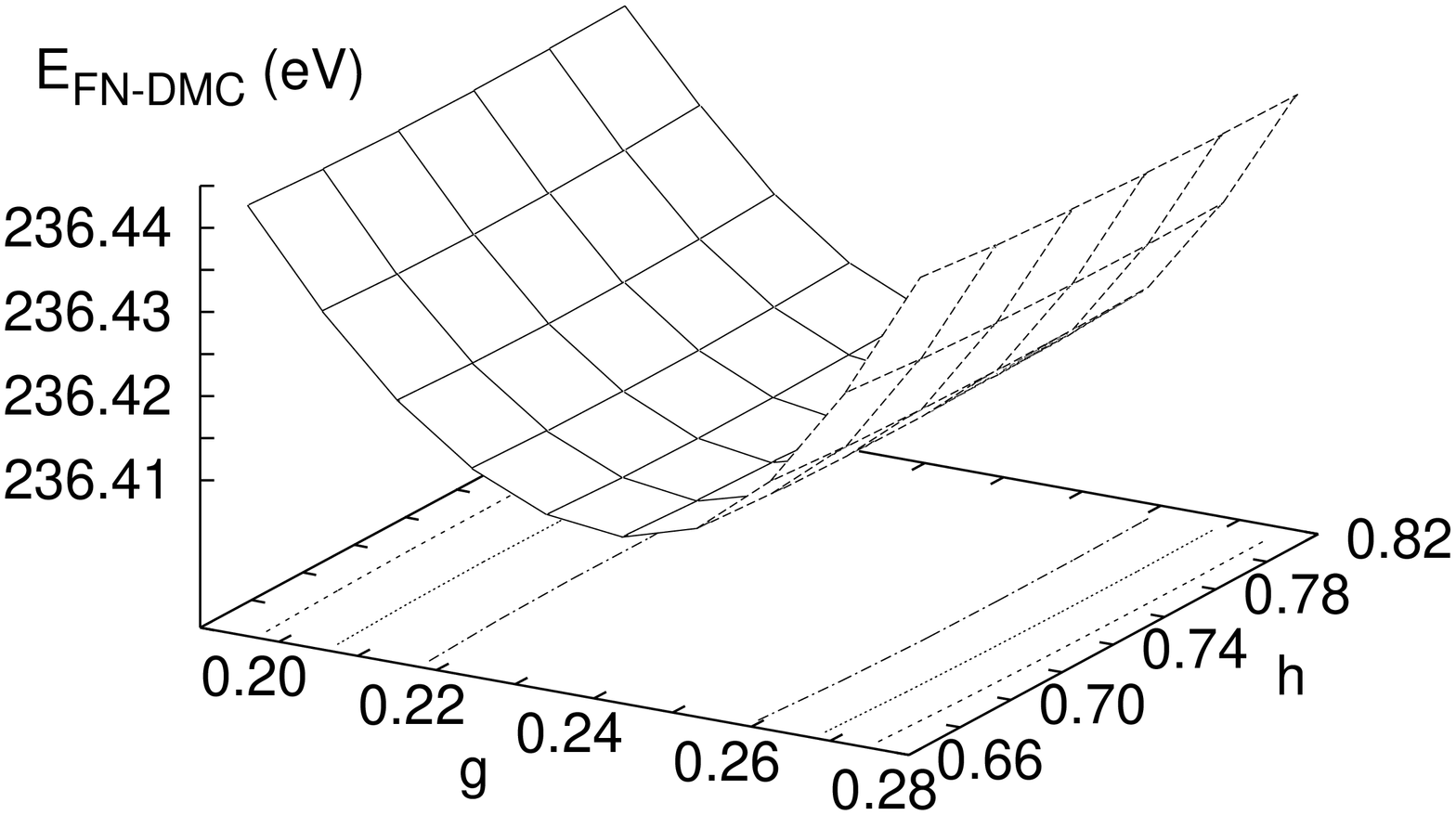}}}}
  \caption[]{\label{screen}
             Correlated sampling for the parameters $g$ and $h$ in
             the generalized Gutzwiller wavefunction 
             $|\Psi_T\rangle= g^D h^{N_c}\,|\Phi\rangle$,
             cf.\ eqn.\ (\ref{scrwf}) in variational (upper plot) and fixed-node
             diffusion Monte Carlo (lower plot). The plots show the energy
             as a function of the Gutzwiller parameters $g$ and $h$, both as
             surfaces and contours. The calculations were done for an fcc
             cluster of 64 molecules with $96+96$ electrons (half-filled
             $t_{1u}$-band), an on-site Hubbard interaction $U=1.25\,eV$, and
             a test charge of $q=1/4$ (in units of the electron charge).
            }
\end{figure}

To study the static dielectric screening for the $t_{1u}$ electrons in the
doped Fullerides,\cite{screenKirchb} we determine the response of the charge 
density to the introduction of a test charge $q$ placed on molecule $i_q$. 
To describe the test charge the term
\begin{equation}
  H_1(q)=qU\sum_{m\sigma} n_{i_q m\sigma} 
\end{equation}
is added to the Hamiltonian (\ref{C60Hamil}). In the spirit of the Gutzwiller
Ansatz we correspondingly add a second Gutzwiller factor to the wavefunction
(\ref{GWFC60}) that reflects the additional interaction term $q U N_{i_q}$:
\begin{equation}\label{scrwf}
  |\Psi_T(g,h)\rangle = g^D h^{N_{i_q}} |\Phi\rangle . 
\end{equation}
Finding the best Gutzwiller parameters is now a two dimensional optimization
problem. Dealing with polynomials in the two variables $g$ and $h$, the
method of correlated sampling works as straightforwardly as described above
for the case of a plain Gutzwiller wavefunction. As an example, 
Fig.\ \ref{screen} shows the result of the optimization, both in variational 
and in fixed-node diffusion Monte Carlo, for a cluster of 64 C$_{60}$ molecules
in an fcc arrangement (periodic boundary conditions) resembling K$_3$C$_{60}$ 
with a test charge $q=1/4$. In practice we first optimize the parameters in
variational Monte Carlo. We then use the optimum VMC parameters as starting 
points for the optimization in the more time consuming fixed-node diffusion
Monte Carlo calculations.

\subsection{Variation of Slater determinant}

In the traditional Gutzwiller Ansatz, the Slater determinant $\Phi$ is the
ground state wavefunction of the non-interacting Hamiltonian. This is,
however, not necessarily the best choice. An alternative would be to use
the Slater determinant $\Phi(U)$ from solving the interacting problem in
Hartree-Fock approximation. We can even interpolate between the two extremes
by doing a Hartree-Fock calculation with a fictitious Hubbard interaction $U_0$
to obtain Slater determinants $\Phi(U_0)$. Yet another family of Slater
determinants $\Phi(H_{\rm stag})$ can be obtained from solving the 
non-interacting Hamiltonian with an added staggered magnetic field, which
lets us control the antiferromagnetic character of the trial function.  
Although optimizing parameters in the Slater determinant cannot be done with 
the method described in the preceeding sections, the cheap optimization of the 
Gutzwiller factors makes it possible to optimize the overall trial function 
without too much effort.

\subsubsection{Staggered magnetic field}

Introducing a staggered magnetic field we can construct Slater determinants
by solving the non-interacting Hamiltonian with an added Zeeman term. To be
specific, we consider K$_3$C$_{60}$ which has a half-filled $t_{1u}$-band.
Since K$_3$C$_{60}$ crystallizes in an fcc lattice, antiferromagnetism is
frustrated and the definition of a staggered magnetic field is not unique.
We split the fcc lattice into two sublattices $A$ and $B$ such that the
frustration is minimized. The Zeeman term is then given by
\begin{equation}
  H_m= H_{\rm stag} \sum_i {\rm sign}(i)\;[n_{i\uparrow}-n_{i\downarrow}]
\end{equation}
with ${\rm sign}(i)=+1$ if $i\in A$ and $-1$ if $i\in B$. It effectively
introduces an on-site energy which, on the same site, has opposite sign for
the two spin orientations, and, for the same spin orientation, has 
opposite sign on the two sublattices. Therefore, hopping to neighboring
sites on different sublattices involves an energy cost of twice the Zeeman
energy. The staggered magnetic field thus not only induces antiferromagnetic 
order in the Slater determinant but also serves to localize the electrons.
This is reflected in the fact that the optimum Gutzwiller parameter is much
larger for Slater determinants constructed from a Hamiltonian with large
$H_{\rm stag}$ than for paramagnetic Slater determinants. 
Varying $H_{\rm stag}$ then interpolates between paramagnetic/itinerant and
antiferromagnetic/localized wavefunctions. 

\begin{figure}
  \centerline{\epsfxsize=3.0in \epsffile{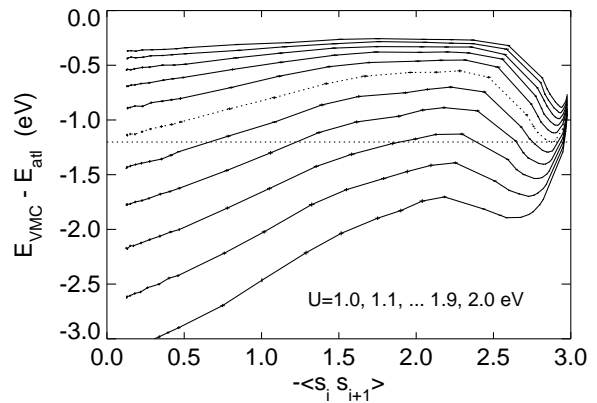}}
  \vspace{2ex}
  \caption[]{\label{stagger_plot}
   Variational energy $E_{\rm VMC}$ for trial functions with different
   character. Plotted are the energies (error-bars, lines are to guide the eye)
   for a Hamiltonian describing K$_3$C$_{60}$ (periodic fcc cluster of 32
   molecules) with Hubbard interaction $U=1.0, 1.1\ldots 1.9, 2.0\,eV$. 
   Instead of the total
   energies $E_{tot}$, we plot the difference of $E_{tot}$ and the energy in
   the atomic limit (each site occupied by three electrons), so that the 
   results for different $U$ can be readily compared.
   The trial functions are of the Gutzwiller type. The Slater determinants
   were determined from diagonalizing the non-interacting Hamiltonian (i.e.\
   setting $U=0$) with a staggered magnetic field $H_{\rm stag}$. This field
   gives rise to an antiferromagnetic correlation of neighboring spins, which 
   is plotted on the abscissa. For $U=1.5\,eV$ (dotted curve) the minima in
   in the paramagnetic and the antiferromagnetic region have about the same
   energy.} 
\end{figure}

The energy expectation values for such trial functions as calculated in
variational Monte Carlo are shown in Fig.\ \ref{stagger_plot}. It shows
$E_{\rm VMC}$ as a function of the antiferromagnetic correlation
\begin{equation}
  \langle s_i s_{i+1} \rangle =
  {1\over N}\sum_{\langle i j\rangle} (n_{i\uparrow}-n_{i\downarrow})\,
                                      (n_{j\uparrow}-n_{j\downarrow}) ,
\end{equation}
where the sum is over the $N$ nearest neighbors. $\langle s_i s_{i+1} \rangle$
is a monotonous function of $H_{\rm stag}$.
For each different value of the Hubbard interaction $U$ we find a curve with
two minima. One minimum is realized for the non-magnetic ($H_{\rm stag}=0$)
trial function. The energy as a function of $U$ scales roughly like
$E_{\rm para}\propto-(1-U/U_c)^2$, as predicted by the Gutzwiller 
approximation. The second minimum is in the antiferromagnetic/localized region
and scales roughly like $E_{AF}\propto -t^2/U$, as expected.
For small $U$ the non-magnetic state is more favorable, while for large $U$
the localized Slater determinant gives lower variational energies. The
crossover is at $U_c\approx1.50\,eV$ (dotted line) and resembles a first
order phase transition.


\subsubsection{Hartree-Fock}

An alternative method for constructing Slater determinants is to use the
interacting Hamiltonian with the \phantom{physical Hubbard interaction}

\vspace{-4ex}
\begin{figure}
  \centerline{\epsfxsize=3.2in \epsffile{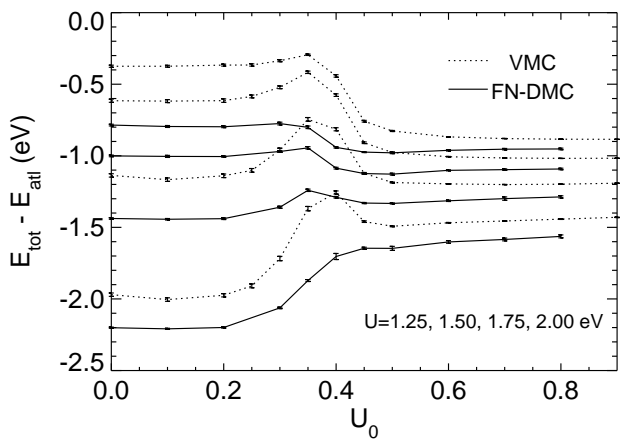}}
  \vspace{2ex}
  \caption[]{\label{U0_plot}
             Dependence of variational (VMC) and fixed-node diffusion
             Monte Carlo (FN-DMC) on the trial function. $U_0$ is the Hubbard
             interaction that was used for the Slater determinant in the
             Gutzwiller wavefunction $\Psi_T(R)=g^{D(R)}\;\Phi(U_0)$.
             The results shown here are the energies (relative to the atomic
             limit) for a Hamiltonian that describes K$_3$C$_{60}$ 
             (32 molecules), with $U$ being varied from $1.25$ (lowest curve) 
             to $2.00\,eV$ (highest curve).}
\end{figure}

\noindent
physical Hubbard interaction $U$ replaced
by a parameter $U_0$ and solve it in Hartree-Fock approximation. In practice
this is done by simply doing an unrestricted self-consistent calculation for
the finite, periodic clusters under consideration, starting from some charge-
and spin-density that breaks the symmetry of the Hamiltonian. It is well 
known that Hartree-Fock favors the antiferromagnetic Mott insulator, 
predicting a Mott transition at much too small values $U_c^{\rm HF}$. 
It is therefore not surprising that good trial functions are obtained for
values of $U_0$ considerably smaller than $U$. For $U_0$ close to zero the
Slater determinant has metallic character, while for somewhat larger $U_0$
there is a metal-insulator transition.
Figure \ref{U0_plot} shows the energy as a function of $U_0$ for the model
of K$_3$C$_{60}$. We find that the results of variational Monte Carlo depend
quite strongly on the parameter $U_0$. As expected, for fixed Hubbard 
interaction $U$ there is a transition from the paramagnetic region for small
$U_0$ to a region where the trial function is antiferromagnetic. 
In fixed-node diffusion Monte Carlo energies are overall lowered and the
dependence on the trial function is much weaker. It seems that here mainly
the character (paramagnetic/antiferromagnetic) of the trial function matters.
For small $U$ trial functions with small $U_0$
give lower energy, while for large $U$ trial functions with larger $U_0$ are
favorable. The crossover coincides with the Mott transition, which takes place
between $U=1.50$ and $1.75\,eV$.\cite{mott}

\section{Summary}

We have presented a convenient and reliable method for optimizing 
Gutzwiller-type trial functions in Monte Carlo calculations. The method is
based on the observation that the expressions for correlated sampling can
be rewritten in terms of polynomials in the Gutzwiller parameters. The method 
can be used both in variational and fixed-node diffusion
Monte Carlo. Because of its reliability and speed it makes optimizing 
Gutzwiller parameters essentially an automatic process. This is especially
convenient when dealing with trial functions that have several Gutzwiller
parameters as in the example on the dielectric screening in K$_3$C$_{60}$.
Given that optimizing the Gutzwiller parameters is quick and easy we can then
focus on optimizing the Slater determinant.

\section*{Acknowledgments}

This work has been supported by the Alexander-von-Humboldt-Stiftung under the
Feodor-Lynen-Program and the Max-Planck-Forschungspreis, and by the Department 
of Energy, grant DEFG 02-96ER45439.

\bibliographystyle{prsty_long}

\begin{thebibliography}{10}
\bibitem{Ntrans}
{This complication arises from the fact that the number of transitions allowed
  by the Pauli principle in general is different for different configurations.
  This issue is discussed in Ref.\ \onlinecite{Horsch}.}
\end{thebibliography}

\end{multicols}
\end{document}